\documentclass[conference]{IEEEtran}
\IEEEoverridecommandlockouts
\usepackage{cite}
\usepackage{amsmath,amssymb,amsfonts}
\usepackage{graphicx}
\usepackage{textcomp}
\usepackage{xcolor}
\usepackage{booktabs}
\usepackage{multirow}
\usepackage{tabularx}
\usepackage{url}
\usepackage{tikz}
\usetikzlibrary{arrows.meta, positioning, fit, backgrounds}

\def\BibTeX{{\rm B\kern-.05em{\sc i\kern-.025em b}\kern-.08em
    T\kern-.1667em\lower.7ex\hbox{E}\kern-.125emX}}

\begin{document}

\title{PISim: A Unified Photonic Inference Simulator for Hardware-Aware Cross-Layer Design-Space Exploration}

\author{\IEEEauthorblockN{Andreas Falkenberg}
\IEEEauthorblockA{\textit{QPI Laboratories, Inc.}\\
Escondido, CA, USA\\
\texttt{andreas@qpilabs.com}}%
\thanks{PISim is available for research evaluation at \protect\url{https://simulator.qpilabs.com}. Access credentials are provided upon request. Project information is available at \protect\url{https://www.qpilabs.com}.}

}

\maketitle

\begin{abstract}
Photonic neural accelerators offer ultra-low latency linear algebra, yet their usable numerical precision, sustainable throughput, and application accuracy remain tightly bounded by surrounding electro-optical interfaces and physical non-idealities. Pre-fabrication design-space exploration (DSE) requires a cross-layer electronic design automation (EDA) framework capable of evaluating disparate optical topologies under identical neural workloads and physical assumptions. We present \textbf{PISim}, a unified Photonic Inference Simulator that couples pretrained deep neural network execution with physics-based models of six accelerator architecture families, six electro-optic modulator technologies, four optical source organizations, and direct or coherent readout schemes. PISim embeds stage-resolved impairment operators---covering control and analog-to-digital quantization, finite extinction ratio, shot noise, correlated relative-intensity noise (RIN), thermal drift, propagation loss, and channel crosstalk---alongside an analytical cycle-level interface timing model to expose electronic-optical bottlenecks. Using simulated MNIST multilayer perceptron (MLP) benchmarks, we demonstrate how PISim identifies optimal precision operating points under multi-impairment stress (peaking at 6 bits for Mach--Zehnder and microring modulators), quantifies a 52.5 percentage-point accuracy gap between VCSEL-array and parallel-fanout laser architectures arising, under the modeled conditions, from power partitioning and correlated source noise, and models interface-limited throughput scaling across emitter-array widths from 32 to 128 channels. This comprehensive framework establishes an extensible, reproducible foundation for accuracy-constrained Pareto optimization prior to photonic tape-out.
\end{abstract}


\section{Introduction}
Photonic integrated circuits (PICs) and free-space optical processors have emerged as compelling platforms for accelerating matrix-vector multiplication (MVM) in deep neural networks (DNNs) \cite{b3, b4, b5, b6, b11}. By exploiting wave propagation, interference, and wavelength-division multiplexing (WDM), optical cores execute linear transformations at the speed of light with minimal dynamic energy dissipation inside the passive optical fabric. However, photonic neural accelerators expose an unusually coupled cross-layer design problem: while the optical core computes a linear transform passively, its effective bit precision, scheduling rate, and end-to-end task accuracy are governed by the surrounding electro-optical system \cite{b11, b12}. Digital-to-analog converters (DACs), electro-optic modulators, optical power distribution networks, photodetectors, transimpedance amplifiers (TIAs), and analog-to-digital converters (ADCs) impose strict bandwidth, dynamic range, and noise constraints that directly impact inference quality.

Consequently, selecting an accelerator topology, modulator technology, source organization, readout scheme, and tile geometry cannot be decoupled from the target neural workload \cite{b12, b13}. Surveys and reviews of photonic matrix multiplication highlight distinct implementation paradigms---including plane-light conversion, coherent interferometer meshes, microring weight banks, and electro-optic multiply--accumulate (MAC) units \cite{b3, b11}. Existing photonic design tools address valuable subsets of this stack: for example, L2ight focuses on scalable mapping, calibration, and in-situ subspace learning for coherent meshes \cite{b1}, LoCI models optical loss and coherent crosstalk in silicon-photonic networks \cite{b2}, and system-level frameworks examine physical-to-algorithm co-exploration and modular scale-out bottlenecks \cite{b12, b13}. Yet architects lack a unified pre-fabrication EDA environment that holds pretrained neural workloads fixed while systematically sweeping across disparate optical architecture families, device transfer curves, correlated noise mechanisms, and electronic interface schedules.

To address this gap, we introduce the \textbf{Photonic Inference Simulator (PISim)}, an integrated cross-layer simulation framework designed for hardware-aware design-space exploration (DSE). PISim enables architects to screen candidate photonic accelerator configurations prior to fabrication while preserving explicit traceability between physical component parameters and application-level inference behavior. Specifically, the contributions of this work are fourfold:
\begin{enumerate}
    \item \textbf{Cross-Architecture Unified Modeling:} A modular simulation engine supporting six photonic accelerator families, six modulator transfer models, four optical source organizations, and both incoherent (direct) and coherent (homodyne) detection paths within a single workload execution contract.
    \item \textbf{Stage-Resolved Physical Impairment Composition:} Mathematical and stochastic formulations of hardware non-idealities---including pre- and post-transfer quantization, finite extinction ratio, signal-dependent shot noise, source-correlated relative-intensity noise (RIN), static fabrication offsets, slow thermal drift, insertion loss, and spatial/spectral crosstalk---injected at their exact physical stages in the forward optical path.
    \item \textbf{Interface-Aware Timing and Bottleneck Accounting:} An analytical cycle-level throughput and latency model separating optical propagation from weight loading, DAC/ADC conversion, integration, and digital post-processing to identify binding electronic-interface bottlenecks across varying array widths and tile counts.
    \item \textbf{Paired Diagnostics and Multi-Objective Exploration:} Repeatable random-stream management enabling paired ideal-versus-photonic confusion matrices, per-class error diagnostics, multi-parameter impairment sweeps, and accuracy-constrained Pareto frontier generation.
\end{enumerate}

The remainder of this extended manuscript is organized as follows. Section~\ref{sec:arch} details the unified simulator architecture, taxonomy of supported photonic topologies, device transfer functions, and source/readout models. Section~\ref{sec:impairments} formalizes the forward optical operators, signed arithmetic encodings, and stage-resolved impairment models. Section~\ref{sec:timing} presents the analytical interface-aware timing, useful-work accounting, and Pareto DSE methodology. Section~\ref{sec:eval} evaluates PISim across five exploratory case studies (E1--E5) examining modulator sensitivity, multi-impairment precision scaling, source organization trade-offs, and modeled array-width throughput scaling. Section~\ref{sec:discussion} discusses validation priorities and roadmap extensions, and Section~\ref{sec:conclusion} concludes the paper.

\section{Unified Architecture and Physical Device Models}
\label{sec:arch}

\subsection{Cross-Layer Workflow and Execution Contract}
PISim bridges high-level deep learning frameworks and low-level photonic physics through a deterministic workload mapping contract. Given a pretrained neural network checkpoint (such as MLPs, ResNet-8, VGG-8, or Mixer-Tiny evaluated on MNIST, Fashion-MNIST, or CIFAR-10) and an input dataset, the simulator constructs a dual-path execution graph. Every input sample $x \in \mathbb{R}^{N_{\text{in}}}$ is propagated simultaneously through an ideal floating-point reference path and a configurable photonic physical path using identical preprocessing and sample ordering.

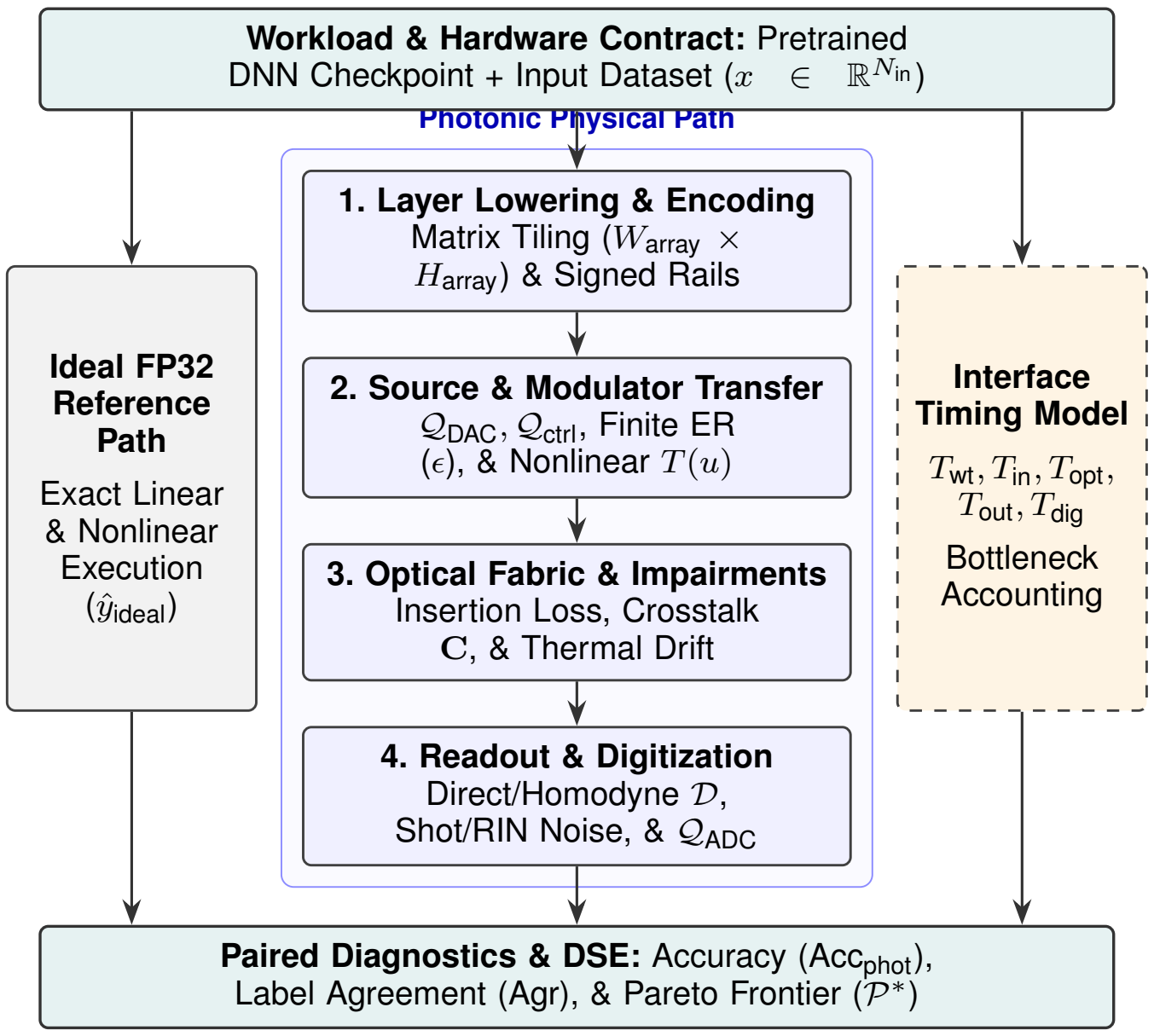
\begin{figure}[t]
\centering
\resizebox{\columnwidth}{!}{%
\begin{tikzpicture}[
    font=\footnotesize\sffamily,
    node distance=0.45cm and 0.45cm,
    block/.style={
        rectangle, draw=black!75, rounded corners=2pt,
        fill=blue!6, text width=4.4cm, align=center,
        minimum height=0.68cm, inner sep=4pt, line width=0.6pt
    },
    refblock/.style={
        rectangle, draw=black!75, rounded corners=2pt,
        fill=gray!10, text width=1.85cm, align=center,
        minimum height=3.8cm, inner sep=4pt, line width=0.6pt
    },
    timeblock/.style={
        rectangle, draw=black!75, dashed, rounded corners=2pt,
        fill=yellow!15!orange!10, text width=1.85cm, align=center,
        minimum height=3.8cm, inner sep=4pt, line width=0.6pt
    },
    ioblock/.style={
        rectangle, draw=black!80, rounded corners=2pt,
        fill=teal!10, text width=8.9cm, align=center,
        minimum height=0.65cm, inner sep=4pt, line width=0.7pt
    },
    arrow/.style={-{Stealth[length=2.2mm, width=1.6mm]}, line width=0.65pt, draw=black!80}
]

\node[ioblock] (input) {\textbf{Workload \& Hardware Contract:} Pretrained DNN Checkpoint + Input Dataset ($x \in \mathbb{R}^{N_{\text{in}}}$)};

\node[block, below=0.5cm of input] (tile) {\textbf{1. Layer Lowering \& Encoding}\\Matrix Tiling ($W_{\text{array}} \times H_{\text{array}}$) \& Signed Rails};
\node[block, below=0.38cm of tile] (mod) {\textbf{2. Source \& Modulator Transfer}\\$\mathcal{Q}_{\text{DAC}}, \mathcal{Q}_{\text{ctrl}}$, Finite ER ($\epsilon$), \& Nonlinear $T(u)$};
\node[block, below=0.38cm of mod] (prop) {\textbf{3. Optical Fabric \& Impairments}\\Insertion Loss, Crosstalk $\mathbf{C}$, \& Thermal Drift};
\node[block, below=0.38cm of prop] (det) {\textbf{4. Readout \& Digitization}\\Direct/Homodyne $\mathcal{D}$, Shot/RIN Noise, \& $\mathcal{Q}_{\text{ADC}}$};

\node[refblock, left=0.38cm of mod.west, anchor=east, yshift=-0.52cm] (ideal) {\textbf{Ideal FP32}\\\textbf{Reference}\\\textbf{Path}\\[4pt]Exact Linear\\\& Nonlinear\\Execution\\($\hat{y}_{\text{ideal}}$)};

\node[timeblock, right=0.38cm of mod.east, anchor=west, yshift=-0.52cm] (timing) {\textbf{Interface}\\\textbf{Timing Model}\\[4pt]$T_{\text{wt}}, T_{\text{in}}, T_{\text{opt}},$\\$T_{\text{out}}, T_{\text{dig}}$\\[4pt]Bottleneck\\Accounting};

\begin{pgfonlayer}{background}
    \node[draw=blue!45, fill=blue!2, rounded corners=3pt, inner sep=5pt,
          fit=(tile)(det), label={[font=\scriptsize\bfseries\sffamily, text=blue!70!black]above:Photonic Physical Path}] (photbox) {};
\end{pgfonlayer}

\node[ioblock, below=0.5cm of det] (output) {\textbf{Paired Diagnostics \& DSE:} Accuracy ($\text{Acc}_{\text{phot}}$), Label Agreement ($\text{Agr}$), \& Pareto Frontier ($\mathcal{P}^*$)};

\draw[arrow] (input.south) -- (tile.north);
\draw[arrow] (input.south -| ideal.north) -- (ideal.north);
\draw[arrow] (input.south -| timing.north) -- (timing.north);

\draw[arrow] (tile.south) -- (mod.north);
\draw[arrow] (mod.south) -- (prop.north);
\draw[arrow] (prop.south) -- (det.north);

\draw[arrow] (det.south) -- (output.north);
\draw[arrow] (ideal.south) -- (ideal.south |- output.north);
\draw[arrow] (timing.south) -- (timing.south |- output.north);

\end{tikzpicture}%
}
\caption{Overview of the PISim unified photonic inference simulation workflow, coupling neural network layer lowering with stage-resolved physical device models, stochastic impairments, and interface timing analysis.}
\label{fig:workflow}
\end{figure}

To prevent ambiguous comparisons between partial optical acceleration and full-system execution, PISim enforces an explicit operator-mapping contract. Linear layers and lowered convolutional layers (via im2col matrix expansion) are partitioned into hardware tiles of dimension $W_{\text{array}} \times H_{\text{array}}$. Non-linear activations, batch normalization, residual additions, and pooling operations are explicitly assigned either to digital electronic post-processing or to optical nonlinear elements (when supported by the chosen topology). For each mapped tile, the simulator tracks:
\begin{itemize}
    \item \textbf{Weight Persistence Policy:} Whether weights are programmed statically once per inference run (e.g., nonvolatile phase-change cells or thermally tuned meshes), reloaded per tile fold, or streamed dynamically per input sample.
    \item \textbf{Active vs. Padded Arithmetic:} Useful MAC operations are counted strictly over active matrix entries, separating zero-padded array channels and redundant multirail encoding lanes from true algorithmic work.
\end{itemize}

\subsection{Supported Photonic Accelerator Families}
Photonic accelerators execute linear transforms using fundamentally different physical field representations. A coherent interferometer mesh manipulates complex optical field amplitudes and phases, whereas an incoherent WDM weight bank sums optical intensities across distinct wavelengths. PISim maintains strict architectural neutrality by decoupling the compute topology from the underlying electro-optic device technology. Table~\ref{tab:architectures} summarizes the six supported accelerator families, their required physical state representations, and their primary EDA exploration variables.

\begin{table*}[t]
\centering
\caption{Taxonomy of Supported Photonic Accelerator Families and EDA Design Variables in PISim}
\label{tab:architectures}
\begin{tabularx}{\textwidth}{>{\raggedright\arraybackslash}p{3.1cm} >{\raggedright\arraybackslash}p{5.3cm} >{\raggedright\arraybackslash}X}
\toprule
\textbf{Architecture Family} & \textbf{Physical Representation \& Forward Operator} & \textbf{Primary EDA Design Variables} \\
\midrule
\textbf{Diffractive Optical Networks} \cite{b6} & Free-space or slab spatial complex field propagation across phase/amplitude diffractive masks onto detector regions. & Mask resolution, layer axial spacing, diffraction efficiency, transverse misalignment, and detector pixel binning. \\
\addlinespace
\textbf{Mach--Zehnder Interferometer (MZI) Meshes} \cite{b4, b7} & Coherent unitary/arbitrary matrix decomposition ($U \Sigma V^\dagger$) via cascaded $2\times 2$ beamsplitters and thermo-/electro-optic phase shifters. & Mesh decomposition topology (e.g., Clements \cite{b7}), phase bit precision, calibration error, and path-dependent insertion loss. \\
\addlinespace
\textbf{Electro-Optic MAC Units} \cite{b5} & Point-to-point or broadcast-and-weight intensity/field modulation followed by spatial or temporal photodetector accumulation. & Modulator transfer curve, signed encoding convention, DAC/ADC sampling rates, and matrix tiling factor. \\
\addlinespace
\textbf{Resonator-Based Inference} \cite{b8} & WDM spectral routing through cascaded microring resonator (MRR) weight banks with differential balanced readout. & Resonance detuning, WDM channel spacing, thermal drift sensitivity, weight state resolution, and inter-channel crosstalk. \\
\addlinespace
\textbf{VCSEL-Array Processors} & Spatially parallel vertical-cavity surface-emitting laser arrays with direct current modulation and optical fan-in/fan-out. & Emitter array width ($N_{\text{vcsel}}$), inter-emitter coupling, nonlinear L-I dynamics, and detector integration window. \\
\addlinespace
\textbf{Free-Space Optical Accelerators} \cite{b9} & Spatial light modulation (SLM/DMD) with cylindrical/Fourier lens optical matrix-vector projection onto camera/photodiode arrays. & Spatial pixel encoding, optical point-spread function, SLM frame refresh rate, and sensor readout schedule. \\
\bottomrule
\end{tabularx}
\end{table*}

\subsection{Electro-Optic Modulator Transfer Models}
PISim parameterizes six distinct electro-optic modulator technologies, mapping normalized digital control drives $u \in$ (or signed drives $u \in [-1, 1]$) to optical field transmission $t(u)$ or intensity transmission $T(u) = |t(u)|^2$. Including finite extinction ratio $\text{ER}_{\text{dB}}$ (where $\epsilon = 10^{-\text{ER}_{\text{dB}}/10}$ denotes the residual transmission floor), the six device models are formulated as follows:

\begin{enumerate}
    \item \textbf{Thin-Film $\text{LiNbO}_3$ Mach--Zehnder Modulator (MZM):} Biased at the quadrature point ($\phi_0 = \pi/2$) with push-pull phase shift $\Delta\phi(u) = \frac{\pi}{2} u$ for $u \in [-1, 1]$, the normalized intensity transmission follows a raised-cosine characteristic:
    \begin{multline}
        T_{\text{MZM}}(u) = \epsilon + (1 - \epsilon) \sin^2\left(\frac{\pi}{4}(1 + u)\right) \\
        = \epsilon + \frac{1 - \epsilon}{2}\left(1 + \sin\left(\frac{\pi}{2}u\right)\right).
        \label{eq:mzm}
    \end{multline}
    In its linearized pre-distorted mode, PISim applies an inverse arcsine lookup $u_{\text{lin}} = \frac{2}{\pi}\arcsin(2\tilde{u} - 1)$ prior to DAC quantization so that DAC discrete levels map linearly onto optical intensity.

    \item \textbf{Silicon Microring Resonator (Ring):} Modeled via an all-pass Lorentzian transfer function governed by resonance wavelength detuning $\delta\lambda(u)$ normalized to the full-width at half-maximum linewidth $\Delta\lambda_{\text{FWHM}}$:
    \begin{equation}
        T_{\text{Ring}}(u) = \epsilon + (1 - \epsilon) \frac{\left(\frac{2(\delta\lambda_0 + \kappa_\lambda u)}{\Delta\lambda_{\text{FWHM}}}\right)^2}{1 + \left(\frac{2(\delta\lambda_0 + \kappa_\lambda u)}{\Delta\lambda_{\text{FWHM}}}\right)^2},
        \label{eq:ring}
    \end{equation}
    where $\delta\lambda_0$ is the static bias offset and $\kappa_\lambda$ is the tuning efficiency. Thermal perturbations shift $\delta\lambda_0$, introducing nonlinear weight distortion.

    \item \textbf{InP Quantum-Confined Stark Effect Electro-Absorption Modulator (InP-QCSE EAM):} Modeled via voltage-dependent exponential optical absorption $\alpha(u)$ over active waveguide length $L_{\text{wg}}$:
    \begin{equation}
        T_{\text{EAM}}(u) = \epsilon + (1 - \epsilon) \exp\left(-\alpha_0 L_{\text{wg}} \left(\frac{u}{u_{\max}}\right)^\gamma\right),
        \label{eq:eam}
    \end{equation}
    with optional phase-amplitude chirp coupling $\Delta\phi_{\text{chirp}}(u) = \frac{\alpha_H}{2} \ln T_{\text{EAM}}(u)$ parameterized by the Henry linewidth enhancement factor $\alpha_H$ when used in coherent paths.

    \item \textbf{Phase-Change Material Cell ($\text{Ge}_2\text{Sb}_2\text{Te}_5$, PCM):} Represents nonvolatile multi-level weights programmed into $M$ discrete crystalline-amorphous conductance/attenuation states $\{s_0, s_1, \dots, s_{M-1}\}$:
    \begin{equation}
        T_{\text{PCM}}(m) = \epsilon + (1 - \epsilon) T_{\text{state}}(s_m), \quad m \in \{0, \dots, M-1\},
        \label{eq:pcm}
    \end{equation}
    capturing non-uniform state spacing and stochastic programming variability without static holding power.

    \item \textbf{Digital Micromirror Device (DLP / DMD):} Operates as a binary spatial reflector $b_k \in \{0, 1\}$ modulated via pulse-width temporal grayscale integration across $B$ bit-planes within integration window $T_{\text{int}}$:
    \begin{equation}
        T_{\text{DMD}}(u) = \epsilon + (1 - \epsilon) \frac{1}{2^B - 1} \sum_{k=0}^{B-1} 2^k b_k(u).
        \label{eq:dmd}
    \end{equation}

    \item \textbf{Surface-Normal QCSE Micro-Pad:} Models vertical-incidence electro-absorption pixels subject to electrical $RC$ bandwidth roll-off and photo-generated carrier screening at high optical intensities:
    \begin{gather}
        T_{\text{micro-pad}}(u, I_{\text{opt}}) = T_{\text{QCSE}}\left(u_{\text{eff}}(t) - \beta_{\text{scr}} I_{\text{opt}}\right), \nonumber \\
        \tau_{RC}\frac{du_{\text{eff}}}{dt} + u_{\text{eff}} = u(t),
        \label{eq:micropad}
    \end{gather}
    where $\tau_{RC}$ is the pad charging time constant and $\beta_{\text{scr}}$ captures saturation due to carrier screening.
\end{enumerate}

\subsection{Optical Source Organizations and Detection Paths}
PISim separates optical source architecture from matrix compute topology across four configurations:
\begin{itemize}
    \item \textbf{VCSEL Array:} $N_{\text{vcsel}}$ independent emitters drive parallel spatial or spectral channels. Each emitter carries independent relative-intensity noise and thermal fluctuations, avoiding $1/N$ passive splitting loss at the expense of per-emitter power and wavelength calibration.
    \item \textbf{Single Laser with Parallel Fanout:} A master continuous-wave (CW) laser of power $P_0$ is split across $N$ channels via a $1\times N$ star coupler or splitter tree. Each channel receives $P_{\text{ch}} = \eta_{\text{split}} P_0 / N$, and master laser RIN propagates as a \emph{spatially correlated} common-mode fluctuation across all $N$ channels simultaneously.
    \item \textbf{Time-Serialized Single Laser:} Full laser power $P_0$ is directed sequentially through a single high-speed modulator-detector channel over $N$ time slots, trading temporal latency for maximum per-sample optical signal-to-noise ratio (SNR) and zero spatial fanout loss.
    \item \textbf{LED Micro-Display:} Incoherent broad-spectrum emitter array restricted strictly to intensity-based direct detection without phase interference.
\end{itemize}

For readout, PISim supports \textbf{direct incoherent detection}, where photodiode current $I_{\text{PD}} = \mathcal{R} P_{\text{opt}}$ is proportional to incident optical power ($\mathcal{R}$ is responsivity in A/W), and \textbf{coherent homodyne detection}, where the signal field $E_{\text{sig}}$ is mixed with a phase-locked local oscillator $E_{\text{LO}}$ in a balanced receiver yielding differential photocurrent $I_{\text{bal}} \propto 2\mathcal{R}\sqrt{P_{\text{sig}} P_{\text{LO}}}\cos(\Delta\theta)$.

\section{Forward Operators and Stage-Resolved Impairments}
\label{sec:impairments}

\subsection{Signed Neural Arithmetic Encoding}
Because neural network weights $W \in \mathbb{R}^{M \times N}$ and activations $x \in \mathbb{R}^N$ take signed values while direct photodetection measures non-negative optical power $P_{\text{opt}} \ge 0$, PISim enforces explicit signed encoding contracts:
\begin{itemize}
    \item \textbf{Differential Dual-Rail Encoding:} Each signed weight $w_{ij}$ is decomposed into positive and negative components $w_{ij} = w_{ij}^+ - w_{ij}^-$ with $w_{ij}^+, w_{ij}^- \ge 0$, routed to balanced photodetector pairs. While doubling hardware rails or time slots, balanced subtraction cancels common-mode intensity offsets and residual extinction floors $\epsilon$.
    \item \textbf{Biased Offset Encoding:} Weights and inputs are shifted to non-negative optical intensities $w_{ij}^{\text{opt}} = w_{ij} + w_{\text{bias}}$ and $x_j^{\text{opt}} = x_j + x_{\text{bias}}$. The accumulated photocurrent contains cross-terms $\sum_j (w_{ij}x_j + w_{ij}x_{\text{bias}} + w_{\text{bias}}x_j + w_{\text{bias}}x_{\text{bias}})$ that are subtracted digitally or via reference channels. Unlike differential encoding, biased offset schemes suffer directly from uncancelled finite extinction floors $\epsilon$.
    \item \textbf{Coherent Field Encoding:} Signed values are encoded directly into optical field amplitude and $\pi$-phase shifts ($E_j \propto x_j$, $W_{ij} \propto w_{ij}$) and recovered via homodyne mixing with an external phase reference \cite{b14}.
\end{itemize}

\subsection{Stage-Resolved Physical Impairment Composition}
Rather than applying an undifferentiated Gaussian perturbation to layer outputs, PISim injects each non-ideality at its exact physical stage along the forward signal chain \cite{b11, b12}:
\begin{equation}
    y_{\text{dig}} = \mathcal{Q}_{\text{ADC}}\Big( \mathcal{D}\Big( \mathcal{H}_{\text{prop}}\big( \mathcal{M}\big( \mathcal{Q}_{\text{DAC}}(x), \mathcal{Q}_{\text{ctrl}}(W) \big) \big) \Big) + n_{\text{det}} \Big),
    \label{eq:chain}
\end{equation}
where $\mathcal{Q}_{\text{DAC}}$ and $\mathcal{Q}_{\text{ctrl}}$ denote input and weight control quantization, $\mathcal{M}(\cdot)$ applies nonlinear modulator transfer and finite extinction ratio, $\mathcal{H}_{\text{prop}}(\cdot)$ applies optical loss, crosstalk, and phase/thermal drift, $\mathcal{D}(\cdot)$ models square-law or homodyne photodetection, $n_{\text{det}}$ injects shot and thermal noise, and $\mathcal{Q}_{\text{ADC}}$ digitizes the electrical readout.

\subsubsection{Multi-Stage Quantization}
PISim separates checkpoint weight precision from $b_{\text{ctrl}}$-bit DAC/modulator control quantization and $b_{\text{ADC}}$-bit receiver digitization \cite{b15}. A normalized control signal $u \in [u_{\min}, u_{\max}]$ is discretized onto $2^{b_{\text{ctrl}}}$ uniform steps:
\begin{equation}
    \mathcal{Q}_{b}(u) = \Delta_b \cdot \text{clip}\left(\left\lfloor \frac{u - u_{\min}}{\Delta_b} + \frac{1}{2} \right\rfloor, 0, 2^b - 1\right) + u_{\min},
    \label{eq:quant}
\end{equation}
where $\Delta_b = (u_{\max} - u_{\min})/(2^b - 1)$. Applying $\mathcal{Q}_{b_{\text{ctrl}}}$ prior to nonlinear device transfer (such as Eq.~\ref{eq:mzm} or Eq.~\ref{eq:ring}) accurately captures non-uniform optical transmission spacing caused by device nonlinearity \cite{b10, b15}.

\subsubsection{Correlated Source RIN and Signal-Dependent Shot Noise}
For a source channel with nominal optical power $P_{\text{src}}$, relative-intensity noise is modeled as a stochastic multiplicative fluctuation $\delta_{\text{RIN}}(t) \sim \mathcal{N}(0, \sigma_{\text{RIN}}^2)$. PISim distinguishes between physical RIN calibration and normalized sensitivity parameters: in physical device mode, $\sigma_{\text{RIN}} = \sqrt{\text{RIN} \cdot B_{\text{eff}}}$ is derived from the physical laser relative-intensity noise power spectral density ($\text{RIN}$ in $\text{Hz}^{-1}$, typically expressed in $\text{dB/Hz}$) integrated over the effective receiver bandwidth $B_{\text{eff}}$, whereas in normalized sensitivity mode, $\sigma_{\text{RIN}}$ is specified directly as a dimensionless fractional standard deviation in normalized simulator control units. Crucially, PISim preserves spatial correlation across source organizations \cite{b10}:
\begin{equation}
    P_{\text{ch}, k}(t) = \frac{\eta_{\text{split}} P_{\text{src}}}{N_{\text{fanout}}} \left(1 + \delta_{\text{RIN}, k}(t)\right),
    \label{eq:rin}
\end{equation}
where $\delta_{\text{RIN}, 1}(t) = \delta_{\text{RIN}, 2}(t) = \dots = \delta_{\text{RIN}, N}(t)$ for a single laser with parallel fanout, whereas $\delta_{\text{RIN}, k}(t)$ are drawn independently across $k \in \{1, \dots, N_{\text{vcsel}}\}$ for a VCSEL array.

At the photodetector, incident optical power $P_{\text{det}, i}$ integrated over window $T_{\text{int}}$ (bandwidth $B = 1/(2T_{\text{int}})$) experiences signal-dependent quantum shot noise:
\begin{equation}
    I_{\text{PD}, i} = \mathcal{R} P_{\text{det}, i} + n_{\text{shot}, i} + n_{\text{th}, i}, \quad \sigma_{\text{shot}, i}^2 = 2 q \mathcal{R} P_{\text{det}, i} B,
    \label{eq:shot}
\end{equation}
where $q$ is the elementary charge and $n_{\text{th}, i}$ represents additive receiver thermal/TIA noise. In normalized sensitivity mode, shot perturbations scale proportionally to $\sigma_{\text{shot}}\sqrt{P_{\text{det}, i}}$.

\subsubsection{Optical Loss, Thermal Drift, and Channel Crosstalk}
Total optical path attenuation $L_{\text{dB}} = L_{\text{split}} + L_{\text{mod}} + L_{\text{prop}} + L_{\text{coup}}$ reduces detected power by $\eta_{\text{opt}} = 10^{-L_{\text{dB}}/10}$, directly degrading SNR in Eq.~(\ref{eq:shot}) \cite{b2, b13}. Thermal drift and static fabrication variations are separated by stochastic lifetime: static fabrication offsets $\Delta\phi_{\text{fab}, i}$ are sampled once per simulated chip instance and held fixed across all dataset samples, whereas slow thermal drift $\Delta\theta_{\text{th}}(t) \sim \mathcal{N}(0, \sigma_{\text{drift}}^2)$ evolves across sample batches, perturbing resonator detuning ($\delta\lambda_0 \leftarrow \delta\lambda_0 + \Delta\lambda_{\text{th}}$) or interferometer phase ($\phi \leftarrow \phi + \Delta\phi_{\text{th}}$) \cite{b10, b14}. Finally, inter-channel crosstalk (from adjacent WDM spectral leakage or waveguide directional coupling) is applied via a sparse coupling matrix $\mathbf{C} \in \mathbb{R}^{N \times N}$ \cite{b2}:
\begin{equation}
    \mathbf{P}_{\text{out}} = \mathbf{C} \, \mathbf{P}_{\text{in}}, \quad C_{ii} = 1 - \sum_{j \neq i} \kappa_{ij}, \quad C_{ij} = \kappa_{ij} \; (i \neq j),
    \label{eq:xtalk}
\end{equation}
where $\kappa_{ij}$ denotes the power (or field) coupling coefficient between channels $i$ and $j$.

\section{Interface-Aware Timing and Design-Space Exploration}
\label{sec:timing}

\subsection{Cycle-Level Throughput and Bottleneck Accounting}
A central pitfall in photonic accelerator evaluation is quoting peak optical propagation throughput ($f_{\text{opt}} = c / (n_{\text{eff}} L_{\text{core}})$) while ignoring electronic conversion and memory feed bottlenecks \cite{b11, b13}. PISim implements an analytical schedule accounting model that decomposes end-to-end inference into five explicit stages for a matrix-vector tile of size $W \times H$ mapped onto an array of physical emitter width $N_w$ and $N_{\text{tiles}}$ parallel tiles \cite{b13, b15}:
\begin{enumerate}
    \item \textbf{Weight Loading ($T_{\text{wt}}$):} Time required to program $W \times H$ weight elements across $N_{\text{prog}}$ programming lanes at update rate $f_{\text{wt}}$, amortized over $N_{\text{reuse}}$ input vectors:
    \begin{equation}
        T_{\text{wt}} = \frac{\lceil (W \cdot H) / N_{\text{prog}} \rceil}{N_{\text{reuse}} \cdot f_{\text{wt}}}.
        \label{eq:twt}
    \end{equation}
    For stationary weights held across an entire batch ($N_{\text{reuse}} \gg 1$), $T_{\text{wt}} \to 0$ \cite{b12}.
    \item \textbf{Input Conversion \& Modulation ($T_{\text{in}}$):} Time to stream $N$ input activations through $N_{\text{DAC}}$ converters operating at sampling rate $f_{\text{DAC}}$ (accounting for $S_{\text{ser}} = N / N_w$ serialization folds):
    \begin{equation}
        T_{\text{in}} = \left\lceil \frac{N}{N_{\text{DAC}}} \right\rceil \frac{1}{f_{\text{DAC}}}.
        \label{eq:tin}
    \end{equation}
    For a time-serialized single laser ($N_{\text{DAC}} = 1, N_w = 1$), $T_{\text{in}}$ scales linearly with vector dimension $N$.
    \item \textbf{Optical Propagation \& Integration ($T_{\text{opt}}$):} Physical time-of-flight $T_{\text{tof}}$ plus photodetector integration window $T_{\text{int}}$ required to meet target SNR:
    \begin{equation}
        T_{\text{opt}} = T_{\text{tof}} + T_{\text{int}}.
        \label{eq:topt}
    \end{equation}
    \item \textbf{Output Digitization ($T_{\text{out}}$):} Time to convert $M$ accumulated output channels across $N_{\text{ADC}}$ receivers at sampling rate $f_{\text{ADC}}$:
    \begin{equation}
        T_{\text{out}} = \left\lceil \frac{M \cdot r_{\text{rail}}}{N_{\text{ADC}}} \right\rceil \frac{1}{f_{\text{ADC}}},
        \label{eq:tout}
    \end{equation}
    where $r_{\text{rail}} \in \{1, 2\}$ accounts for single-ended versus differential dual-rail readout.
    \item \textbf{Digital Post-Processing ($T_{\text{dig}}$):} Partial-sum accumulation across tiles, bias subtraction, scaling, and nonlinear activation execution in the digital host/ASIC \cite{b15}.
\end{enumerate}

Under pipelined streaming execution, the amortized per-sample latency $T_{\text{sample}}$ and sustainable inference throughput $\Theta_{\text{inf}}$ (in inferences/s) are bounded by the slowest stage:
\begin{equation}
    T_{\text{sample}} = \max\left(T_{\text{wt}}, \, T_{\text{in}}, \, T_{\text{opt}}, \, T_{\text{out}}, \, T_{\text{dig}}\right), \quad \Theta_{\text{inf}} = \frac{1}{T_{\text{sample}}},
    \label{eq:throughput}
\end{equation}
whereas single-request unpipelined latency equals the serial sum $T_{\text{lat}} = T_{\text{wt}} + T_{\text{in}} + T_{\text{opt}} + T_{\text{out}} + T_{\text{dig}}$. Useful arithmetic throughput (in MAC/s or OP/s, with $1\text{ MAC} = 2\text{ OPs}$) is computed strictly over active algorithmic operations $\text{MAC}_{\text{useful}}$:
\begin{equation}
    \Theta_{\text{OP}} = \frac{2 \cdot \text{MAC}_{\text{useful}}}{T_{\text{sample}}}.
    \label{eq:tops}
\end{equation}
By inspecting which term dominates Eq.~(\ref{eq:throughput}), PISim automatically flags whether a candidate design is \emph{DAC-bound}, \emph{serialization-bound}, \emph{integration/SNR-bound}, or \emph{ADC-bound} \cite{b11}.

\subsection{Paired Reproducibility and Pareto Optimization}
To improve statistical comparability across design sweeps, PISim partitions pseudo-random number generator (PRNG) streams by physical mechanism and stochastic lifetime:
\begin{multline}
    \mathcal{S}_{\text{run}} = \Big(\mathcal{S}_{\text{fab}}(\text{chip\_id}), \; \mathcal{S}_{\text{drift}}(\text{batch\_id}), \\
    \mathcal{S}_{\text{fast}}(\text{sample\_id}, \text{layer\_id})\Big).
    \label{eq:seeds}
\end{multline}
Because every configuration evaluates identical input samples with matched stream partitioning, PISim computes both aggregate top-1 accuracy $\text{Acc}_{\text{phot}}$ and \emph{paired label agreement} $\text{Agr} = \frac{1}{K}\sum_{k=1}^K \mathbb{I}(\hat{y}_{\text{phot}}^{(k)} = \hat{y}_{\text{ideal}}^{(k)})$, alongside paired confusion matrices that expose class-specific bias induced by extinction floors or drift \cite{b10}.

For multi-objective DSE, let $\mathcal{X}$ denote the discrete-continuous configuration space spanning architecture family, modulator device, source organization, bit precision $b$, array width $N_w$, and tile count $N_{\text{tiles}}$ \cite{b11, b12}. Given an application accuracy constraint $\text{Acc}_{\text{phot}}(\mathbf{x}) \ge \text{Acc}_{\min}$ and a cost/latency objective vector $\mathbf{f}(\mathbf{x}) = \big(T_{\text{sample}}(\mathbf{x}), \, N_{\text{det}}(\mathbf{x}), \, C_{\text{proxy}}(\mathbf{x})\big)$, PISim extracts the Pareto-optimal frontier $\mathcal{P}^*$:
\begin{multline}
    \mathcal{P}^* = \Big\{ \mathbf{x} \in \mathcal{X}_{\text{feas}} \;\Big|\; \nexists \mathbf{x}' \in \mathcal{X}_{\text{feas}} \text{ s.t. } \\
    \mathbf{f}(\mathbf{x}') \preceq \mathbf{f}(\mathbf{x}) \land \mathbf{f}(\mathbf{x}') \neq \mathbf{f}(\mathbf{x}) \Big\}.
    \label{eq:pareto}
\end{multline}

\section{Experimental Evaluation and DSE Case Studies}
\label{sec:eval}

\subsection{Experimental Setup}
Although PISim implements the broader model taxonomy described in Section~\ref{sec:arch}, the experiments reported here exercise a representative subset selected to isolate device-, source-, precision-, and interface-scaling effects. Specifically, we evaluate PISim using a pretrained 3-layer MNIST MLP ($784 \rightarrow 64 \rightarrow 10$, 50,816 weight bytes) across 1,000 test samples (indices 0--999) with deterministic PRNG seed 33. The ideal floating-point reference checkpoint achieves \textbf{97.1\%} classification accuracy on this evaluation split. Table~\ref{tab:config} summarizes the documented experimental and physical simulation parameters across all case studies. Across five structured experiments (E1--E5), we investigate modulator equivalence under ideal baselines, precision scaling under multi-impairment stress, optical source organization trade-offs, and modeled array-width throughput scaling.

\begin{table}[t]
\centering
\caption{Default Simulation and Experimental Configuration}
\label{tab:config}
\footnotesize
\setlength{\tabcolsep}{4pt}
\begin{tabularx}{\columnwidth}{@{}>{\raggedright\arraybackslash}p{2.95cm} >{\raggedright\arraybackslash}X@{}}
\toprule
\textbf{Parameter} & \textbf{Documented Value / Setting} \\
\midrule
Dataset & MNIST (test split, indices 0--999) \\
Sample Count & 1,000 samples \\
Network Topology & 3-layer MLP ($784 \rightarrow 64 \rightarrow 10$; 50,816 weight bytes) \\
Random Seed & 33 (deterministic PRNG) \\
Reference Accuracy & 97.1\% (ideal FP32) \\
Array Width ($N_w$) & $\{32, 64, 128\}$ channels (E5) \\
DAC / Control Precision & $\{4, 6, 8\}$ bits (E2--E3); 8 bits (E4) \\
Optical Source & Single-laser parallel fanout (E1--E3); VCSEL, serialized (E4--E5) \\
Source Power ($P_{\text{src}}$) & 1.0 (normalized, E2--E3) \\
Detection Mode & Incoherent direct detection (E2--E4) \\
Laser RIN ($\sigma_{\text{RIN}}$) & 0 (E1--E2); 0.01 (normalized, E3--E4) \\
Shot Noise ($\sigma_{\text{shot}}$) & 0 (E1--E2); 0.01 (normalized, E3--E4) \\
Thermal Drift ($\sigma_{\text{drift}}$) & 0 (E1--E2); 0.01 (normalized, E3--E4) \\
Extinction Ratio ($\text{ER}$) & 30 dB ($\epsilon = 10^{-3}$, E3--E4) \\
Insertion Loss ($L_{\text{ins}}$) & 0 dB (E1--E2); 3 dB (E3--E4) \\
\bottomrule
\end{tabularx}
\end{table}

\subsection{E1: Clean-Reference Modulator Comparison}
In Experiment E1, we compare three distinct electro-optic modulator technologies---Thin-Film $\text{LiNbO}_3$ MZM, Silicon Microring Resonator (Ring), and Surface-Normal QCSE Micro-Pad---under clean operating conditions (all stochastic noise, drift, loss, and crosstalk controls disabled) using single-laser parallel fanout. All three modulator architectures achieve \textbf{97.1\%} photonic inference accuracy, matching the 97.1\% ideal floating-point reference with zero degradation. This confirms that when calibrated to their linear operating regimes without physical impairments, distinct modulator transfer functions preserve full network fidelity.

\subsection{E2--E3: Precision Sweeps Under Clean and Multi-Impairment Stress}
To quantify how control bit precision interacts with hardware non-idealities, Experiments E2 and E3 sweep \texttt{quant\_bits} $\in \{4, 6, 8\}$ across MZM, Ring, and QCSE configurations using single-laser parallel fanout, incoherent detection, and normalized source power $P_{\text{src}} = 1.0$:
\begin{itemize}
    \item \textbf{E2 (Clean Quantization Baseline):} All five physical impairment controls are disabled.
    \item \textbf{E3 (Combined Multi-Impairment Stress):} Five simultaneous non-idealities are enabled in normalized sensitivity mode: shot noise ($\sigma_{\text{shot}} = 0.01$), laser RIN ($\sigma_{\text{RIN}} = 0.01$), thermal drift ($\sigma_{\text{drift}} = 0.01$), finite extinction ratio ($\text{ER} = 30\text{ dB}$, $\epsilon = 10^{-3}$), and optical insertion loss ($L_{\text{ins}} = 3\text{ dB}$). Here, normalized parameters such as $\sigma_{\text{RIN}} = 0.01$ (as well as $\sigma_{\text{shot}} = 0.01$ and $\sigma_{\text{drift}} = 0.01$) are controlled, dimensionless simulation sensitivity parameters and should not be interpreted as calibrated physical measurements from a specific laser or photonic device. Crosstalk, nonlinear optical effects, speckle, and phase noise remain disabled.
\end{itemize}

\begin{table}[t]
\centering
\caption{E2--E3 Per-Bit Photonic Inference Accuracy (\%) on MNIST MLP (Ideal Reference = 97.1\%)}
\label{tab:e2e3}
\begin{tabular}{ccccc}
\toprule
\textbf{Precision} & \textbf{E2 Clean} & \textbf{E3 Stress} & \textbf{E3 Stress} & \textbf{E3 Stress} \\
\textbf{(Bits)} & \textbf{(All Three)} & \textbf{MZM} & \textbf{Ring} & \textbf{QCSE} \\
\midrule
4 bits & 95.9\% & 35.7\% & 35.7\% & 24.6\% \\
6 bits & 97.0\% & \textbf{46.4\%} & \textbf{46.4\%} & 15.9\% \\
8 bits & \textbf{97.1\%} & 44.5\% & 44.5\% & \textbf{19.0\%} \\
\bottomrule
\end{tabular}
\end{table}

Table~\ref{tab:e2e3} summarizes the numerical accuracy across precision levels, revealing two critical EDA insights:
\begin{enumerate}
    \item \textbf{Clean Precision Saturation (E2):} Under impairment-free conditions, all three modulators exhibit identical monotonic accuracy recovery: 95.9\% at 4 bits ($-1.2\%$ drop), 97.0\% at 6 bits ($-0.1\%$ drop), and full recovery to 97.1\% at 8 bits. Thus, 6-bit control precision suffices for nearly lossless representation in the absence of analog noise.
    \item \textbf{Non-Monotonic Precision Behavior Under Stress (E3):} Under heavy combined noise ($\sigma = 0.01$) and 30 dB extinction ratio, accuracy degrades severely across all modulators, and scaling to higher bit-widths no longer guarantees monotonic improvement \cite{b10}. Both MZM and Ring peak at \textbf{46.4\%} at 6 bits before dropping slightly to 44.5\% at 8 bits, as finer quantization steps fall below the analog noise floor ($\Delta_8 \approx 0.0039 < \sigma_{\text{noise}}$) and capture stochastic perturbations without improving signal fidelity. QCSE exhibits substantially higher sensitivity under identical stress settings, peaking at 24.6\% at 4 bits and dropping to 15.9\% at 6 bits and 19.0\% at 8 bits due to its distinct transfer sensitivity.
\end{enumerate}

\subsection{E4: Impact of Optical Source Organization Under Stress}
Experiment E4 isolates the architectural impact of optical source organization by holding the workload (MNIST MLP), modulator (MZM), detection mode (incoherent), precision (8 bits), and combined E3 stress profile ($\sigma_{\text{shot}} = \sigma_{\text{RIN}} = \sigma_{\text{drift}} = 0.01$, $\text{ER} = 30\text{ dB}$, $L_{\text{ins}} = 3\text{ dB}$) completely fixed while varying only the optical source topology.

\begin{figure}[t]
\centering
\includegraphics[width=\columnwidth]{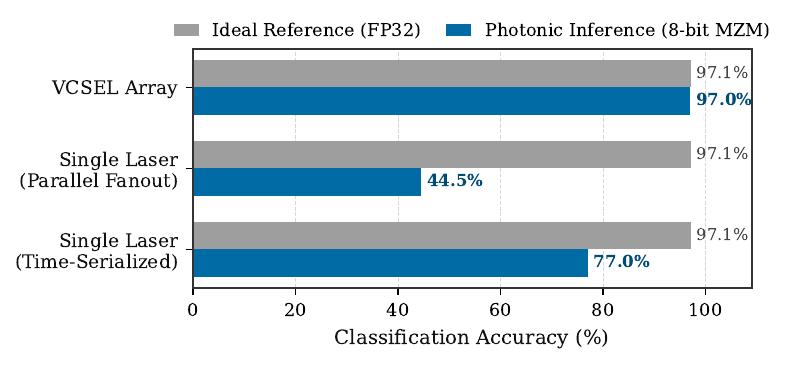}
\caption{Photonic inference accuracy across optical source organizations under matched 8-bit MZM multi-impairment stress (E4). Eliminating $1/N$ fanout splitting loss and spatially correlated RIN restores accuracy from 44.5\% to 97.0\%.}
\label{fig:e4_source}
\end{figure}

As shown in Fig.~\ref{fig:e4_source}, source organization produces dramatic accuracy divergence despite identical device parameters:
\begin{itemize}
    \item \textbf{VCSEL Array (\textbf{97.0\%}):} Achieves near-ideal accuracy (within 0.1\% of the 97.1\% reference) even under full E3 stress. Because each channel is driven by a dedicated emitter without passive $1/N$ optical splitting loss and without spatially correlated common-mode RIN across the entire layer vector, per-channel SNR remains well above the noise floor.
    \item \textbf{Single-Laser Time-Serialized (\textbf{77.0\%}):} Outperforms parallel fanout by $+32.5\%$ because full master laser power $P_0$ is delivered to a single active channel per time slot without $1/N$ power division, mitigating shot-noise degradation while still experiencing temporal RIN and drift across sequential steps.
    \item \textbf{Single-Laser Parallel Fanout (\textbf{44.5\%}):} Suffers an accuracy penalty of 52.5 percentage points relative to the VCSEL array arising, under the modeled simulation conditions, from the combined effect of $1/N$ optical power partitioning (reducing detected signal power relative to shot noise and extinction floor) and spatially correlated RIN perturbing all vector elements coherently.
\end{itemize}
This finding underscores a core cross-layer EDA trade-off: parallel fanout from a single laser simplifies emitter packaging but requires substantially higher launch power or differential noise cancellation to match VCSEL-array robustness.

\subsection{E5: Array-Width Throughput and Latency Scaling}
Experiment E5 evaluates analytically modeled interface-limited system throughput across emitter-array widths $N_w \in \{32, 64, 128\}$ for the MNIST MLP (50,816 bytes weight memory) across VCSEL/MZM, single-laser parallel/MZM, and single-laser serialized/MZM configurations. Table~\ref{tab:e5} reports modeled inference throughput (in million inferences/s) and amortized per-sample latency at peak throughput.

\begin{table}[t]
\centering
\caption{E5 Modeled Inference Throughput (Million Inferences/s) and Peak Amortized Latency Across Array Widths}
\label{tab:e5}
\setlength{\tabcolsep}{4pt}
\begin{tabular}{cccc}
\toprule
\textbf{Array Width} & \textbf{VCSEL /} & \textbf{Parallel} & \textbf{Serialized} \\
\textbf{($N_w$)} & \textbf{MZM} & \textbf{Laser / MZM} & \textbf{Laser / MZM} \\
\midrule
32 & 5.563 M/s & 5.563 M/s & \textbf{12.399 M/s} \\
64 & 10.710 M/s & 10.710 M/s & \textbf{12.496 M/s} \\
128 & \textbf{19.584 M/s} & \textbf{19.584 M/s} & 12.545 M/s \\
\midrule
\textbf{Min Latency} & \textbf{51.06 ns} & \textbf{51.06 ns} & 79.71 ns \\
\bottomrule
\end{tabular}
\end{table}

The modeled scaling curves in Table~\ref{tab:e5} expose a clear architectural crossover governed by electronic-interface scheduling:
\begin{itemize}
    \item \textbf{Spatial Parallel Scaling (VCSEL \& Parallel Laser):} Both parallel architectures exhibit identical modeled throughput, scaling nearly linearly with array width in the analytical schedule model from 5.563 M inferences/s at $N_w = 32$ to 10.710 M/s at $N_w = 64$ ($1.93\times$) and 19.584 M/s at $N_w = 128$ ($3.52\times$), reaching a modeled amortized per-sample latency of \textbf{51.06 ns}. Increasing $N_w$ reduces the number of sequential matrix-vector folding cycles required to process the 784-element input layer.
    \item \textbf{Time-Serialized Saturation:} The serialized single-laser configuration maintains nearly flat modeled throughput across array widths (12.399 M/s at $N_w = 32$ to 12.545 M/s at $N_w = 128$, with a modeled peak amortized latency of \textbf{79.71 ns}) because simulated execution is bottlenecked by single-channel sequential modulation/digitization rather than spatial emitter width.
    \item \textbf{Crossover and Joint Accuracy-Throughput Selection:} At narrow array widths ($N_w \in \{32, 64\}$), the high-speed serialized laser achieves $1.17\times$ to $2.23\times$ higher modeled throughput than spatial arrays. However, at $N_w = 128$, parallel arrays surpass the serialized bottleneck in the timing model by $1.56\times$. Combining E4 and E5 reveals that while Parallel Laser/MZM and VCSEL/MZM achieve identical modeled peak throughput (19.584 M/s at $N_w = 128$), only \textbf{VCSEL/MZM} simultaneously satisfies high accuracy constraints (97.0\% vs. 44.5\% under E3 stress), positioning it on the simulated accuracy-constrained Pareto frontier.
\end{itemize}

\section{Discussion and Roadmap for Hardware Validation}
\label{sec:discussion}
The exploratory evaluations in Section~\ref{sec:eval} demonstrate how PISim exposes non-obvious interactions between device physics, noise correlation, and interface scheduling prior to fabrication. Follow-on multi-dimensional perturbation studies have already extended these device models to deeper CIFAR-10 convolutional workloads (a compact four-layer CNN and ResNet-8), mapping two-dimensional RIN-versus-drift accuracy response surfaces for Mach--Zehnder and microring matrix multipliers \cite{b10}. To transition from normalized pre-fabrication screening to silicon-calibrated sign-off, three extensions are prioritized:
\begin{enumerate}
    \item \textbf{Physical Parameter Calibration and Multi-Seed Confidence:} Replacing normalized simulator sigma controls with foundry-calibrated PDK distributions (linking $\sigma_{\text{RIN}}$, $\tau_{RC}$, $\beta_{\text{scr}}$, and thermal drift directly to physical $\text{dB/Hz}$ and Kelvin metrics) \cite{b11} and running multi-seed Monte Carlo ensembles to establish statistical confidence intervals on Pareto frontiers.
    \item \textbf{Joint Accuracy-Throughput-Energy Sweeps (E6):} Extending E5 to evaluate matched paired accuracy at every array width $N_w$ and tile count $N_{\text{tiles}}$ across multi-core scale-out topologies \cite{b12, b13}, while incorporating a validated system-level energy accounting boundary covering laser wall-plug efficiency, modulator drivers, thermal heaters, TIAs, and DAC/ADC conversion energy \cite{b11, b15}.
    \item \textbf{Experimental Silicon Correlation:} Benchmarking PISim forward predictions against fabricated photonic test chips across deeper vision and sequence workloads (ResNet-8, VGG-8, and Mixer-Tiny) \cite{b10, b12}.
\end{enumerate}

\section{Conclusion}
\label{sec:conclusion}
We presented the \textbf{Photonic Inference Simulator (PISim)}, a unified electronic design automation framework for pre-fabrication design-space exploration of photonic neural accelerators. By integrating six optical architecture families, six parameterized electro-optic modulator transfer models, four source organizations, and stage-resolved physical impairment operators with an analytical interface timing model, PISim enables architects to evaluate end-to-end neural network accuracy and system bottlenecks on equal footing. Case studies on MNIST inference reveal critical cross-layer design rules: precision scaling under multi-impairment stress exhibits non-monotonic optima where finer quantization falls below analog noise floors; VCSEL arrays preserve 97.0\% accuracy under stress profiles that degrade single-laser parallel fanout to 44.5\%; and, under the analytical interface timing model, spatial parallel arrays overtake time-serialized lasers in throughput at width $N_w = 128$ (19.58 M inferences/s, 51.06 ns modeled latency). PISim provides a rigorous, reproducible foundation for accuracy-constrained photonic accelerator co-design.

\end{document}